\documentstyle[aps,preprint]{revtex}
\input {amssym.def}
\input {amssym.tex}
\everymath{\rm}
\everydisplay{\rm}

\begin{document}
\title{Only Fermi-Liquids are Metals}
\author{C. M. Varma}
\address{Bell Laboratories, Lucent Technologies \\
Murray Hill, NJ  07974}
\maketitle

\begin{abstract}
Any singular deviation from Landau Fermi-liquid theory 
appears to lead, for arbitrarily small
concentration of impurities coupling to a non-conserved quantity, to a vanishing density
of states at the chemical potential and infinite resistivity as temperature approaches
zero.  Applications to copper-oxide metals including the temperature dependence of the
anisotropy in resistivity, and to other cases of non Fermi-liquids are discussed.
\end{abstract}

\newpage
The resistivity in the c-direction, $\rho_c (T)$, in the normal phase of most
copper-oxide (CuO) compounds increases with decreasing temperature while the inplane
resistivity $\rho_{a,b} (T)$ has the opposite behavior, which for compositions near those
for the highest $T_c$ is proportional to $T$ down to $T \simeq T_c$.$^1$  One may be led to
suppose that, if superconductivity were not to intervene, $\rho_c \rightarrow \infty$
while $\rho_{a,b} \rightarrow$ finite value as $T \rightarrow 0$.  For any finite
quantum-mechanical transfer matrix element $t_\perp$ between adjacent planes, the
asymptotic low temperature dependence in different directions must be identical for 
$T << T_{xa}$, where
\begin{equation}
\tau^{-1}_{in} \left ( T_{xa} \right ) \simeq t_\perp  .
\end{equation}
Here $\tau^{-1}_{in} (T)$ is the inelastic scattering rate.  Anderson and Zhou$^2$
conjectured that the renormalized matrix elements
$t_\perp (T) \rightarrow 0$ in CuO compounds at low temperatures due to
orthogonality effects.  An alternative conjecture$^3$ put forward to resolve the issue is
that $\rho_{a,b}$ also $\rightarrow \infty$ as $T \rightarrow 0$ due to impurity
scattering in a non Fermi-liquid.  Recent experiments$^{4,5}$ measuring the resistivity
at low temperatures by suppressing $T_c$ in a large magnetic field support this
conjecture and find $\rho_{a,b} (T) \sim \rho_c \sim \ell n T$ at low temperature.
Similar behavior is also found (without applying a magnetic field) in samples of the
single layer Bi compound.$^6$  Here theoretical support for the conjecture that the
resistivity of a non-Fermi-liquid is infinity for $T \rightarrow 0$ for any finite
concentration of impurities as well as the logarithm temperature dependence are obtained.

A Landau Fermi-liquid has the property that the real part of the
single particle self-energy 
\begin{equation}
Re \: \Sigma \: ( \omega , T, k_F ) \sim x^{\alpha} , \mbox{with}\: \alpha = 1 .
\end{equation}
Here $x = \omega$ for $| \omega | >> T$ and $= \pi T$ for $T >> | \omega |$.
For $\alpha < 1$, the pole in the single particle Green's function, the quasiparticle,
vanishes at the chemical potential and is replaced by a branch-cut.  $\alpha < 1$ may be
used to characterize a non Fermi-liquid.  The gentlest departure from a Landau
Fermi-liquid with
\begin{equation}
Re \: \Sigma \: ( \omega , T, k_F ) \sim \omega \: \left| \ell n 
\frac{\omega_c}{x} \right|
\end{equation}
has been termed a marginal Fermi-liquid.$^7$

The observed linear temperature dependence of the resistivity 
and corresponding behavior of the frequency dependent conductivity
implies directly that the
momentum scattering rate, $\tau_{mom}^{-1}$, in Cu-O compounds is proportional to 
$max ( | \omega | , T )$.  The imaginary part of the single particle 
self-energy $Im \Sigma ( \omega , T, k_F )$ cannot have a higher
power dependence on $( | \omega |, T )$ than the momentum scattering rate.  
Through Kramers-Kronig transformation, this implies $\alpha < 1$.  The marginal case,
Eq. (3) is consistent with the measured tunneling conductance$^8$ as
well as the deduced electronic heat capacity,$^9$ as is the low temperature resistivity
derived here.  

I discuss below the $( \omega , T )$
dependence of the scattering rate from {\it s-wave} scattering off impurities for the
non-Fermi-liquids characterized in the pure limit by the marginal case as
well as other $\alpha < 1$.  This can be done to a considerable extent without reference to
any microscopic theory of the non-Fermi-liquid.  Also, for a given $\alpha$, the
dimensionality $d$ will not be important as long as $d > 1$.  Of course $\alpha$ may
depend on $d$.

Consider impurities which in the absence of electron-electron interactions have finite
s-wave scattering amplitude.$^{10}$  The vertex correction due to the interactions will in
general introduce scattering into higher partial waves.  But there is no reason why the
s-wave scattering should vanish.  Consider only this part for which the forward
scattering limit characterizes the properties.  In this limit, a Ward identity provides
the vertex including the effect of electron-electron interactions.

Let the bare impurity potential be
\begin{equation}
V_{imp} (r) = \sum_i v \: \delta ( r - r_i )  ,
\end{equation}
where $\left\{ r_i \right\}$ are the position of the impurities assumed randomly distributed.
We will assume also that $v / \epsilon_F << 1$ and the concentration of 
impurities $c$ is low
enough so that the mean-free path $\ell_0$ calculated without electronic
renormalizations satisfies $(k_F \ell_0 )^{-1} << 1$.  The s-wave part of the
renormalized scattering from a given impurity is
\begin{equation}
\tilde{v} = v \: lim_{q \rightarrow 0} \: \Lambda_{{k_F}, \omega} \: (q, 0)
\end{equation}
where $\Lambda_{{k_F}, \omega} (q, 0)$ is the irreducible vertex due to electron-electron
interactions for elastic scattering with momentum $q$.

If an impurity couples to a non-conserved quantity, (for instance for Cu-O 
compounds, it can change the
equilibrium charge difference between Cu and O in a unit cell or alter the local kinetic
energy in a Cu-O bond) a Ward-identity gives that$^{11,12,13}$ (at $T = 0$),
\begin{equation}
\lim_{q \rightarrow 0} \: \Lambda_{{k_F}, \omega} \: (q, 0) = z^{-1} ( \omega )
\end{equation}
where
\begin{equation}
z^{-1} ( \omega ) = \left( 1 -
\frac{\partial}{\partial \omega} \:Re \: \Sigma \: (k_F , \omega ) \right) .
\end{equation}
The self-energy due to impurities from such a vertex for a given impurity configuration
illustrated in Fig. (1a) is
\begin{equation}
\Sigma_{imp} ( \omega ) =
\frac{\tilde{V}_{imp} ( \omega )}{1 - \tilde{V}_{imp} ( \omega ) 
{\displaystyle \sum_k} G ({\bf k},
\omega )}
\end{equation}
where $\tilde{V}_{imp} ( \omega ) = \sum_i \tilde{v} ( \omega ) \delta (r - Ri)$ and 
$G ( {\bf k}, \omega )$ is the exact single particle Green's function including the
effect of impurities.

In Eq. (8), vertex renormalization at a given impurity alone has been considered.
Vertex renormalization which spans across impurities, such as in Fig. (1b) are
proportional to $( \omega / \epsilon_F )$ so that (8) may be considered asymptotically exact.

The strategy adopted here is to first evaluate a subset of the processes in the
expression (8), which are expected to be the most singular.  The result so obtained is
used to calculate corrections which
show that the expectation is borne out.  The most singular processes for
the impurity averaged self-energy are expected to be given by the self-consistent Born
approximation represented by Fig. (1c):
\begin{equation}
Im \: \Sigma^0_{imp} ( \omega ) = c \left(
\frac{v}{z} \right)^2 Im \sum_k G^{\prime} ({\bf k}, \omega )  ,
\end{equation}
where we include in $G^{\prime} ({\bf k}, \omega )$ the self-energy 
$\Sigma^{\prime}$ which includes $\Sigma ( \omega , k )$ as
well as $\Sigma^0_{imp} ( \omega )$.

Assuming a constant density of states $N (0)$ over a bandwidth $\approx 2 \epsilon_F$ in
the pure limit, (other assumptions about the pure density of states do not produce any
essential difference), Eq. (9) gives
\begin{equation}
Im \: \Sigma^0_{imp} ( \omega ) = \left(
\frac{1}{2 \tau_0} \right) \frac{1}{z^2} \frac{2}{\pi}
\mbox{tan}^{-1} \left(
\frac{\epsilon_F}{Im \Sigma^{\prime} (\omega )} \right)  ,
\end{equation}
where $\tau_0 = \ell_0 / v_F$.  For $z$ evaluated from the marginal self-energy and for
$\epsilon_F / Im \: \Sigma^0_{imp} (\omega ) >> 1$,
\begin{equation}
Im \: \Sigma^0_{imp} ( \omega ) \approx
\frac{1}{2 \tau_0} \left(1 + \lambda \ell n \: \frac{\omega_c}{\omega_o} \right)^2 .
\end{equation}

In the opposite limit
\begin{equation}
Im \: \Sigma^0_{imp} ( \omega ) \approx \lambda
\sqrt{\frac{\epsilon_F}{2 \tau_o}} \: \ell n \left(
\frac{\omega_c}{\omega} \right)  .
\end{equation}
The above results are valid for $\omega >> T$.  For $\omega << T ,\: \pi T$ 
should be substituted
for $\omega$ in Eqs. (10-11).  The cross-over from (11) to (12) occurs for
\begin{equation}
\left( \ell n \frac{\omega_c}{max ( \omega_x , \pi T_x )} \right)^2
\approx \frac{2 \epsilon_F \tau_0}{\lambda^2}
\end{equation}
or at $\omega = 0$ at
\begin{equation}
T_x \approx \frac{\omega_c}{\pi} \: exp \left( - \lambda^{-1} \sqrt{k_F \ell_0} \right)  .
\end{equation}
Note that below $\omega_x $ the density of states near the Fermi-energy:
\begin{equation}
\frac{1}{\pi} \: Im \: \sum_k G^{\prime} (k, \omega ) = N (0) / \ell n \:
( \omega_c / \omega )  ,
\end{equation}
which approaches zero at the chemical potential.  

The two major omissions in evaluating
(7) in the self-consistent Born approximation are (i) possible strong renormalized
scattering from a given impurity and (ii) interference effects between scattering at
different impurities.  Consider the former.  For Fermi-liquids the singular scattering
from a given impurity in the Born-approximation is remedied by evaluating the t-matrix,
Fig. (1d):
\begin{equation}
\tilde{t}_i = \frac{v_i /z}{1 - v_i \: z^{-1} \: {\displaystyle \sum_k} \: G (k, \omega )}
\end{equation}
The $\tilde {t}$ matrix obeys the unitary limit $\tilde{t} \: N (0) \rightarrow 1$ as $z$
becomes very small (but not zero), as in heavy Fermions.  This is because for Fermi-liquids
${\displaystyle \sum_k} G (k, 0) = N (0)$, the bare density of states,
independent of the value of $z$.

We may check the validity of the self-consistent Born approximation by evaluating the
corrections to the self-energy due to (16) by using $G = G^{\prime}$ in (16).  Now we
note that on impurity averaging, one gets correction terms
\begin{equation}
Im \: \Sigma_{imp} (\omega ) = Im  \Sigma^o_{imp} ( \omega )
\left[ 1 + \left( \frac{v}{z} \sum_k G^{\prime} (k, \omega ) \right)^2 +
\left( \frac{v}{z} \sum_k G^{\prime} (k, \omega ) \right)^4 + ... \right]
\end{equation}
Using (15), we see that the singularity due to $z^{-1}$ is cancelled out to all orders 
leaving an analytic correction of $0 ( v N(0) ) << 1$.

We may similarly consider correction due to crossed graphs, Fig. (1e).  Again the
renormalized density of state cancels the singularity in $v/z$ and the corrections are
successive powers of $c v N(0)$.  Since interference between scatterers is not
important, the wavefunctions are not localized; the important effect is the vanishing of
the density of states at the chemical potential, as in Eq. (15).

For $\alpha < 1$, Eq. (12) is modified to
\begin{equation}
Im \: \Sigma^o_{imp} ( \omega ) \approx \lambda \sqrt{\frac{\epsilon_F}{2 \tau_0}}
\: \left( \frac{\omega_c}{\omega} \right)^{1 - \alpha} ,
\end{equation}
so the effects of impurities are more singular.

The most important effect neglected above is the renormalization by impurities of
the fluctuation which produced the non-Fermi-liquid state in the pure limit.  This
cannot be discussed without a microscopic theory for such fluctuations.  
A theory of such fluctuations has been constructed$^9$.  The corrections to the
fluctuations due to impurities have been briefly examined which appear not to change the
results here in an essential way.  This matter is however far from being settled.  At
this point it is best to leave this as an assumption and to point out that the results
thus obtained appear to agree with experiments, as discussed below.

A prediction following from Eq. (15) is that the specific heat $C (T)$ and the magnetic
susceptibility $\chi (T)$ have the forms 
$C (T) / T \sim \chi (T) \sim ( \ell n T )^{-1}$, for $T << T_x$.  For s-wave scattering
by impurities, there are no backward scattering (vertex) corrections for the
calculation of resistivity.  The resistivity can then be calculated from the single
particle Green's function alone and is proportional to 
$\ell n \frac{\omega_c}{T}$ for $T << T_x$.

Consider now the anisotropic resistivity of quasi-two-dimensional materials.  Two
distinct temperature scales need to be defined to discuss the anisotropic resistivity of
non-Fermi-liquids.  One of them is $T_{x,a}$ defined in Eq. (1), and the other is $T_x$
defined in Eq. (14).  $T_{x,a}$ is enough to discuss Fermi-liquids.  For $T >> T_{x,a}$, the
layers are mutually phase incoherent and momentum in is not conserved
in the process of electron transfer between the planes.  One can calculate
$\rho^{-1}_c$ by a tunneling rate calculation.  The tunneling rate (if the tunneling
matrix element has some momentum dependence) is proportional to the (in-plane) inelastic
scattering rate.  Then $\rho^{-1}_c (T) \sim \rho_{a,b} (T)$.  For $T << T_{x,a}$ a
coherent propagation must prevail in a Fermi-liquid and $\rho_c (T) \sim \rho_{a,b} (T)$.
This behavior is indeed observed, for example in $Sr_2 Ru O_4$.$^{14}$ 

For non-Fermi-liquids two limiting cases can easily be distinguished.  (i) $T_{x,a} <<
T_x$:  which occurs for highly anisotropic and fairly dirty materials.  The resistivity
in the c-direction increases with decreasing temperature for all temperatures but below
$T_{x,a}$.  Both $\rho_{a,b}$ and $\rho_c$ have similar temperature dependences below
$T_{x,a}$ but for $T >> T_{x,a} , \: \rho^{-1}_c \sim \rho_{a,b}$.  Very high quality
samples of $Y Ba_2 Cu_3 O_{6.9}$ appear to fall in this class if we assume that in this
low anisotropy material $T_{x,a}$ is above the measured temperature range. The more
anisotropic dirty materials are consistent with class (i) but in the high quality
samples studied $T_x$ is quite low, not too far from $T_{x,a}$.
For instance, in the measurements of Boebinger et al.$^5$ on $La_{1.85} Sr_{.15} Cu O_4$,
$k_F \ell_o$ is estimated to be about 15 which with 
$\omega_c \approx 2 \times 10^3 K$ and $\lambda \approx 1$, estimated from the slope of
the linear restivity and the optical conductivity$^7$ gives $T_x \approx 0 (20 K)$ from 
Eq. (17).  This is consistent with the temperature at which $\rho_{a,b} (T)$ has a minima in
the experiments and asymptotically below which a logarithmic temperature dependence is
observed.  Systematic estimations of $k_F \ell_o$ are not yet available to test 
Eq. (14).  

The results here have possible applications to other situations where interactions lead
to non-Fermi-liquid properties in the pure limit.  These include the quantum critical
points in itinerant ferromagnets and antiferromagnets as well as the mysterious
transitions in $Ce Cu_{6-x} Au_x$.$^{15}$  It would appear from the results here that disorder
is strongly relevant at metal-insulator transitions otherwise driven by electronic
correlations.  The effect of disorder in driving $N (0) \rightarrow 0$ appears stronger
than the localization effect which sets in only for $k_F \ell_o \sim O (1)$.
The pseudo-particle Green's function at the $\nu = \frac{1}{2}$ quantum
Hall effect also have a marginal Fermi-liquid form.$^{16}$  But since in this case the Green's
function is a gauge dependent object, a separate investigation is required for physical
quantities.

I wish to thank Y. Ando, G. Boebinger, G. Kotliar, M. R. Li, S. Martin, and K. Miyake
for useful discussions.
\newpage
\section*{REFERENCES}
\begin{enumerate}
\item 
See for example B. Batlogg in {\it High Temperature Superconductivity}.  The Los Alamos
Symposium - 1989, edited by K. S. Bedell et al. (Addison Wesley, Reading, MA; 1989).
\item
P. W. Anderson and Z. Zhou, Phys. Rev. Lett. {\bf 60}, 2557 (1988).
\item
G. Kotliar and C. M. Varma, Physica {\bf A 167}, 288 (1990);
G. Kotliar et al., Europhysics Lett. {\bf 63}, 1996 (1989).
\item
Y. Ando et al., Phys. Rev. Lett. {\bf 75}, 4662 (1995).
\item
G. Boebinger et al., Phys. Rev. Lett, {\bf 77}, (1996).
\item
S. Martin and C. M. Varma, unpublished analysis of results in S. Martin et al.,
Phys. Rev. B {\bf41}, 846 (1990).
\item
C. M. Varma et al., Phys. Rev. Lett. {\bf 63}, 1996 (1989).
\item
P. B. Littlewood and C. M. Varma, Phys. Rev. B. {\bf 45}, 12636 (1992).
\item
C. M. Varma, Phys. Rev. (submitted), available on http://www.lanl.gov
as cond-mat/9607105. 
\item
See for example, H. Ikeda and K. Miyake, J. Phys. Soc. Japan {\bf 6}, (1996).
\item
O. Betbeder-Matibet and P. Nozieres, Annals of Physics, {\bf 37}, 17 (1966).
\item
J. S. Langer, Phys. Rev. {\bf 128}, 110 (1962).
\item
If impurities couple to conserved quantities such as {\it total} charge or spin density
(in the absence of spin-orbit scattering), Eq. (6) is multiplied by 
$(1 + F^s_0 )^{-1}$ or $( 1 + F^a_o )^{-1}$ where $F^{s,a}_0$ are the lowest
spin-symmetric and anti-symmetric Landau parameters.  Impurity scattering rates can then
be singular only if compressibility or spin-susceptibility in the pure system diverges.
We do not consider such a situation here.
\item
Y. Maeno et al., Nature {\bf372}, 532 (1994).
\item
H. von Lohneysen, J. Phys. Condens. Matt. {\bf 8}, 9689 (1996)
\item
B. I. Halperin, P. A. Lee and N. Read, Phys. Rev. B {\bf 47}, 7312 (1993).
\end{enumerate}
\newpage
\section*{FIGURE CAPTIONS}
\begin{itemize}
\item[Fig. 1]
Graphs describing the calculation of the impurity self-energy:  
\\
\noindent
(a) The impurity
self-energy for a given configuration (i, j, ...) of impurities.  The line with arrows
is the single particle propagator including renormalization due to electron-electron
interaction and the circle connected to the dashed lines is the impurity vertex
renormalized for the effect of interactions.  
\\
\noindent
(b) Vertex corrections due to interactions neglected in (a).  The wavy line represents
electron-electron interactions.
\\
\noindent
(c) Impurity self-energy (after configuration averaging) in the self-consistent Born
approximation.  The thick line includes the self-energy due to interactions as well as
due to impurities.
\\
\noindent
(d) The $t$-matrix with renormalized vertices of interaction from an impurity at site
$i$.
\\
\noindent
(e) The crossed-graphs for interference effects between two impurities at site $i$ and
at site $j$. 
\end{itemize}
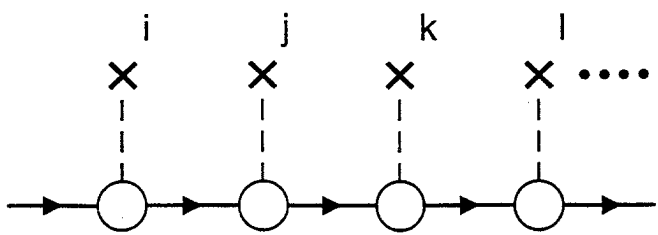
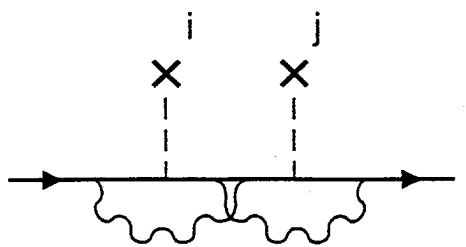
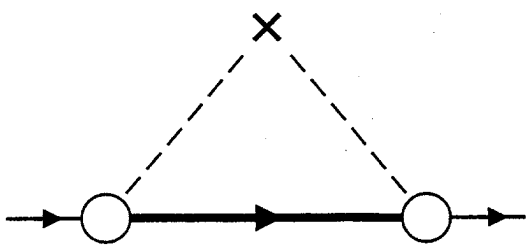
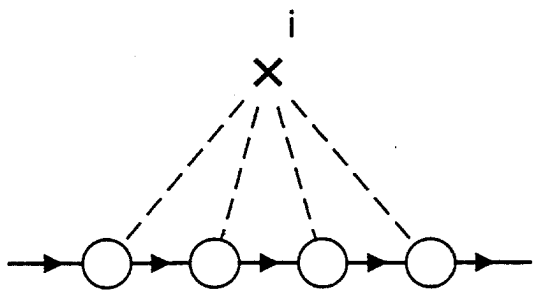
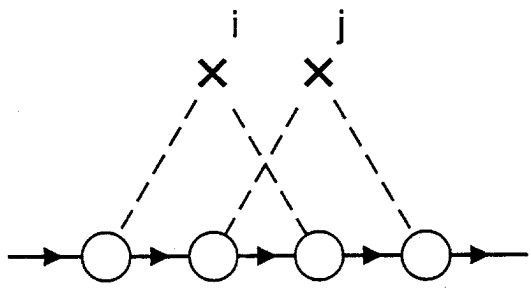
\end{document}